# Cell-Permeable Tumor Suppressor Peptides for Cancer Therapy: Back to the Future

**Razvan Tudor Radulescu**

*Molecular Concepts Research (MCR), Munich, Germany*

E-mail: ratura@gmx.net

**Abstract**

Miniaturization is a hallmark of modern technologies. Notably, this feature has not spared molecular biology and its potential applications. Towards developing more effective therapeutics against cancer, studies began to explore more than a decade ago how natural tumor suppression could be translated into antineoplastic drugs. To this end, investigators focused on major constituents of a central pathway that protects cells against neoplastic transformation: the nuclear retinoblastoma protein (RB) pathway. As such, peptide mimetics of RB, p16 and p21 were developed. Likewise, the p53 and von Hippel-Lindau gene products which affect indirectly the RB pathway provided additional templates for the development of anti-proliferative peptides. Each of the peptides derived from these distinct tumor suppressors was made cell-permeable by its ligation to an amino acid sequence conferring cellular internalization. Details reviewed here reveal that through the application of such anti-cancer peptide therapeutics alone or in conjunction whenever synergy is to expect, the dark era of chemotherapy will likely be overcome, at last.

It is interesting that major insights within various scientific fields often display common features. The "helix" idea is a classic that has emerged both in the early studies by Pauling *et al.* dealing with protein structure and those subsequently investigating the nature of the gene. Another example that connects even more disparate areas is that concerning miniaturization. This principle of down-sizing while upholding or even surpassing the capabilities of a larger template- initially mainly reserved to the discipline of electronics, particularly to the design of ever smaller central processing units e.g. of those driving personal computers, ultimately resulting in tiny microchips- has meanwhile reached molecular biology, specifically the development of preclinical therapeutics.

In this context, studies conducted more than a decade ago suggested that one could compress a given protein function into a much shorter peptide sequence. This idea then served as a basis for bold theoretical leaps according to which one could combine several such sequences, each embodying a different function, to create fusion polypeptides that, despite being much smaller than a functionally comparable protein, could substitute or even be superior to their parent molecules. One such avenue was outlined by synthetic inducible biological response amplifiers (SIBRAs), multi-functional polypeptides proposed to become activated upon appropriate cues and characterized by several features such as ligand binding, cellular internalization/nuclear translocation as well as DNA binding and trancriptional activation, ultimately ensuring an appropriate host response to the cue and gradually disappearing once the cue had vanished (1).

Yet, it was soon realized that accomplishing SIBRAs may be a long way to go and, hence, the feasibility and efficacy of less complex synthetic peptides was addressed as a preliminary stage towards SIBRAs. It was at about the same time that the existence of a retinoblastoma protein (RB) pathway of central importance to tumor suppression and therefore inactivated in most human tumors was clearly established (2). This pathway mainly involves RB (or yet a relative thereof such as p107 or p130) and three types of proteins: a cyclin, a cyclin-dependent kinase (cdk) and an inhibitor of such kinase such as p16 and p21. Since cancer posed a formidable treatment challenge then as it still does now, the question arose whether the tumor-suppressive components of the RB pathway, i.e. RB, p16 and p21, could be imitated by much smaller artificial peptides which would thus be more suitable to be applied as drugs than the larger natural proteins serving as their templates.

As far as human RB- a 928-amino acid protein- is concerned, coupling a 6-amino acid RB fragment with binding affinity for the LXCXE motif in growth-promoting proteins such as insulin and the E7 protein of human papilloma virus (HPV) 16 to a cellular internalization sequence- more precisely, a nuclear localization sequence-, resulted in a fusion polypeptide capable of inhibiting cell cycle progression (3,4) and, upon further modifications conferring protease resistance, cytotoxic to various human tumor cells *in vitro* (5). It was further elucidated that the more precise mechanisms of action of these RB-derived peptides are an interference with insulin-RB dimerization (6,7) and the induction of highly elevated levels of intracellular p21 (8). Most intriguingly, however, is these peptides´ demonstrated *in vivo* antineoplastic activity against syngeneic mouse tumors (9) and human lung cancer xenografts in nude mice (10,11).

As a variation on the RB peptide theme, it has more recently been shown that a 39-amino acid peptide sequence present in the RB family protein pRb2/p130, if joined to an internalization sequence, yields a fusion polypeptide with *in vitro* and *in vivo* anti-tumor properties (12).

Furthermore, proteins upstream from RB were equally explored towards the generation of anti-cancer peptide mimetics. Accordingly, a p16 peptide was designed, joined to a cell-penetrating sequence and such fusion polypeptide then shown to block the cell cycle progression of RB-positive cancer cells *in vitro*, yet not that of RB-negative malignant cells (13), reflecting the necessity for functional RB to be present if p16 (or a peptide derived therefrom) is to exert growth-inhibitory effects. This observation has also been confirmed in an *in vivo* experimental mouse model for human melanoma (14). Moreover, the cdk inhibitor p21 could also be translated into an active antineoplastic peptide (15).

Last but not least, tumor suppressors such as p53 and the VHL gene product, respectively, which are known to employ RB as a downstream effector to interfere with oncogenesis equally served as templates for the design of anti-cancer peptides (16-18).

However, after the "Golden Age" of cell-permeable tumor suppressor peptides between 1997 and 2003, this promising direction could not keep up the same fast pace mainly because of two reasons.

Firstly, tyrosine kinases as drug targets had captured the attention of most anti-cancer drug developers and their sponsors around the turn of millennium. Yet, despite the introduction of Bcr-Abl and HER2 inhibitors into the clinical therapy of cancer, a genuine breakthrough, especially in treating metastasis, is still beyond reach for such (non-peptide) drugs.

In fact, as previously outlined within the framework of a cybernetics of cancer signalling (19), therapeutic blockade of (cell surface) tyrosine kinases is particularly prone to inducing resistance phenomena since such proteins are highly redundant for a cancer cell given that, in terms of signal transduction, they are located considerably upstream from the nucleocrine (20) and RB (2) pathways. This preceding view has just recently been confirmed once again by the necessity to employ a "cocktail" of several tyrosine kinase inhibitors in order to achieve antineoplastic effects (21). However, it has to be emphasized that, besides the peril of adverse drug interactions potentially arising from any combination regimen, even such "cocktails" will predictably entrain the emergence of significant resistance, in this case specifically involving cytoplasmic proteins relaying (cell surface) tyrosine kinases to the key players in the nucleus, for instance by selecting for mutant PI3kinases displaying constitutive (oncogenic) activity (22).

By contrast, interventions aimed at directly influencing the nucleocrine and RB pathways- located one step away from gene expression and thus representing a signal transduction level of minimal redundancy and maximal entropy, hence making resistance development rather unlikely- should represent a most promising therapeutic approach. It could be accomplished by cell-permeable RB pathway-derived tumor suppressor peptides entering not only cancer cells, but also non-malignant cells.

This fundamentally novel proposal (23) to affect both types of cells in the same way- thus contradicting the current dogma of targeting just malignant cells while sparing normal cells- is crucial since such approach would globally counteract oncoprotein-driven metastasis and consequently prevent that neoplastic transformation proceeds at the (epi)genetic level in otherwise morphologically normal cells, for instance in normal cells adjacent to a breast carcinoma (24).

Secondly, maintaining a swift development rate for nuclear tumor suppressor-derived peptide therapeutics was further hampered by the long-standing and widespread fallacy according to which peptides *per se* are allegedly not good drugs.

However, starting with Vincent du Vigneaud´s success story on the synthesis of oxytocin, a peptide to be later administered as a drug for the induction of labor, back in the 1950s until today when the T20 peptide proved to be an efficient therapeutic against HIV infections, peptides have repeatedly underscored their great potential for direct usage as drugs. Yet, as often in the history of medicine, even the best treatment principles need individuals who champion their use, whether it was penicillin or are peptide drugs now, particularly given the latter´s exciting prospects one can certainly envisage (25).

If cancer is to be effectively treated with cell-penetrating tumor suppressor peptides in the future, then, besides the option for single drug treatment, e.g. of RB-negative malignancies by means of RB-like peptides, one should also explore potential synergies between various such peptides. For instance, in the case of patients with RB-positive tumors, it would be interesting to investigate whether a potential efficacy of p16 peptides can be augmented by the addition of RB-derived peptides in an attempt to increase the quantity of functional RB proteins *in situ* to mediate p16 peptide-initiated antineoplastic effects. The ultimate challenge will be to achieve synergy between distinct tumor suppressor peptides at the expense of minimal or no adverse interactions between such candidate drugs while minimizing the emergence of resistance to therapy by combining several such peptide drugs. Though not a simple task, the journey back into the bright horizon of cell-permeable tumor suppressor peptides should be worth completing.

## Acknowledgments

I would like to thank Judah Folkman and Daniel von Hoff for the encouragement as well as support they have given me along my anti-cancer peptide research avenue.